\documentclass[showpacs,preprintnumbers,onecolumn]{revtex4}%
\usepackage{amsfonts}
\usepackage{amsmath}
\usepackage{graphicx}
\usepackage{dcolumn}
\usepackage{bm}
\usepackage{amssymb}%
\providecommand{\U}[1]{\protect\rule{.1in}{.1in}}
\begin{document}
\preprint{ }
\title{Search for a dark matter particle family}
\author{Yukio Tomozawa}
\affiliation{Michigan Center for Theoretical Physics, Randall Laboratory of Physics,
University of Michigan, Ann Arbor, MI. 48109-1040}
\date{\today }

\begin{abstract}
I suggest a simple signature for new particles which are unstable partners of
a dark matter particle. The suggested mass range is from 8 TeV to 3 PeV,\ the
former being the mass of the dark matter particle and the latter being the
knee energy mass scale from the cosmic ray energy spectrum. It can be the
energy spectrum of a specific particle such as a muon, a neutrino, jets or any
other particles produced in cosmic ray showers, as long as the spectrum is
measued. As for the detection of a 3 PeV particle by the neutrino energy
spectrum, all dark matter targets throughout the galaxy that are bombarded by
high energy cosmic rays and high energy dark matter particles contribute to
the process. This is new in the study of dark matter physics.

\end{abstract}

\maketitle

\section{Introduction}

LHC experiments are underway with the hope that new particles in a dark matter
family might be observed. So far no new partitcles have been discovered. From
high energy gamma ray searches the prognosis for the existence of a relatively
low mass (less than 1 TeV) dark matter particle (DMP) seems unlikely.
HESS\cite{hess1} concluded that there is no gamma ray peak below 2 TeV, while
above 2 TeV up to 10 TeV, there is gamma ray excess above the power law
extension from the lower energy data. This suggests that the discovery of new
particles in a DMP family might require various kinds of cosmic ray detectors,
as was the case for the discovery of strange particles in the pre-1950 era, a
golden age of cosmic ray physics. All new particles were found exclusively in
cosmic ray detectors, before the arrival of the accelerator era. In the next
section, I summarize the expected mass range for a DMP family to see what we
are dealing with. The method which we develop for detection is of more general
application, however.

\section{Mass range for DMP species}

In the standard model, all particles are unstable except the particles of
lowest mass, protons, electrons and neutrinos. Similarly, a supersymmetric
theory suggests that all particles of a family are unstable except for the
lowest mass state, in this case the DMP. There may be multiple DMP's,
counterparts of protons, electrons and neutrinos in the standard model. In
order to specify a possible mass range for DMP partners, I discuss a theory
where the mass scale has been explicitly derived from observational data. In a
series of articles\cite{cr1}-\cite{cr9} since 1985, the author has presented a
model for the emission of high energy particles from AGN. The following is a
summary of the model.

1) Quantum effects on gravity yield repulsive forces at short
distances\cite{cr1},\cite{cr3}.

2) The collapse of black holes results in explosive bounce back motion with
the emission of high energy particles.

3) Consideration of the Penrose diagram eliminates the horizon problem for
black holes\cite{cr4}. Black holes are not black any more.

4) The knee energy for high energy cosmic rays can be understood as a split
between a radiation-dominated expansion and a matter-dominated expansion, not
unlike that in the expansion of the universe. (See page 10 of the lecture
notes\cite{cr1}-\cite{cr3}.)

5) Neutrinos and gamma rays as well as cosmic rays should have the same
spectral index for each AGN. They should show a knee energy phenomenon, a
break in the energy spectral index at 3 PeV, similar to that for the cosmic
ray energy spectrum.

6) The recent announcement by Hawking rescinding an earlier claim about the
information paradox\cite{hawking} is consistent with this model.

It is worthwhile to mention that this model has been supported by recent data
from the Pierre Auger Observatory, which has found a possible correlation
between the sources of high energy cosmic rays and AGN.

Further discussion of the knee energy in the model yields the existence of a
new mass scale in the knee energy range, in order to have the knee energy
phenomenon in the cosmic ray spectrum\cite{crnew}. The following are
additional features of the model.

7) The proposed new particle with mass in the knee energy range (at 3 PeV) may
not be stable, as in the case of the standard model. The standard model has
particles at the 100 GeV mass scale, such as W and Z bosons, which are
unstable. If it is a member of a supersymmetric multiplet and weakly
interacting with ordinary particles, the stable particle of lowest mass
becomes a candidate for a DMP. The only requirement is that such particles
must be present in AGN or black holes so that the phenomenon of the knee
energy is observed when cosmic rays are emitted from AGN.

8) Using the supersymmetric theory of GLMR-RS (Giudice-Luty-Murayama-Rattazzi;
Randall-Sundrum)\cite{glmr}, \cite{rs}, the lowest mass corresponding to a
knee energy mass of 3 PeV is 8 TeV. It is shown that the sum of 8 gamma ray
observations from unknown sources\cite{hess2} has a definite peak at 7.6 $\pm$
0.1 TeV\cite{hessanal1}, \cite{hessanal2}.

9) There are several other particles with mass between 8 TeV \ and 3 PeV in
the GLMR-RS theory.

We assume that the target mass range for the search is from 8 TeV to 3 Pev. In
particular, the 3 PeV target is of prime importance, since it is a starting
point for the discovery of a new mass scale. Moreover, it provides bases for
matter-dominated expansion of black holes and possibly of the universe. It is
expected that such particles are produced abundantly in the process of cosmic
ray production from AGN as well as in the process of universe expansion. But
most particles produced in AGN decay. They are produced in pairs in high
energy cosmic ray showers and subsequently decay. In this article, we aim to
find a signature for such particles.

One may call the new particle system the Cion system. This is an acronym for
Cosmic Interface Particle. It is also taken from the Chinese word for knee, Xi
(pronounced as shi). The particle at 3 PeV may be called a prime-Cion and the
DMP particle at 8 TeV a dm-Cion.

\section{Jet production in high energy collisions}

It is quite common to observe many jets in a high energy collision. Jets seem
more frequent than would be dictated by phase space considerations. In other
words, a high energy collision produces a relatively small number of jet
entities rather than producing many particles which share the momenta.
Ignoring small momenta pependicular to the jet direction, a jet may be
approximated as a single particle. By the same token, the decay of a massive
particle may be approximated by a relatively fewer number of jet or particle
decays. In such a situation, significant probability is shared by the smallest
number of particles, i. e., two body decay. If one assumes two body decays,
one finds particles carrying half the energy of the parent mass in the rest
frame of a parent particle.

The next simplest decay mode is three body decay. The phase space for three
body decay has a triangular form peaked at the highest energy, which is half
the mass of the parent particle. This is generally true unless the matrix
element vanishes at the highest energy. In that case, the peak shifts to a
lower energy.

\section{Mass spectra of the GLMR-RS supersymmetric theory}

Since the knee energy of the cosmic ray energy spectrum implies a new mass
scale in nature, according to the theory proposed by the present author in
1985, and the model predicted a correlation between high energy cosmic rays
and AGN that was observed by the Pierre Auger Observatory\cite{auger}, it
seems natural to expect new physics at a mass scale of 3 PeV. Then one would
look for a supersymmetric theory which has a big mass ratio, so that a
relatively low mass for a dark matter particle can be predicted from a
practical observational point of view. From that consideration, the author
came to the conclusion that the analysis of the GLMR-RS theory\cite{glmr},
\cite{rs}, \cite{pevss} was the most appropriate. In that theory, the basic
mass relations are given by the gaugino mass parameters,%
\begin{align}
M_{1}  &  =\frac{11\alpha}{4\pi\cos^{2}\theta_{W}}m_{3/2}=8.9\ast
10^{-3}m_{3/2},\\
M_{2}  &  =\frac{\alpha}{4\pi\sin^{2}\theta_{W}}m_{3/2}=2.7\ast10^{-3}%
m_{3/2},\\
M_{3}  &  =-\frac{3\alpha_{s}}{4\pi}m_{3/2}=-2.6\ast10^{-2}m_{3/2}%
\end{align}
before loop corrections, where $\alpha$, $\alpha_{s}$ and $\theta_{W}$ are the
fine structure constant, strong coupling constant and the weak interaction
angle respectively.

With the assumption for the largest mass,
\begin{equation}
m_{3/2}=3PeV,
\end{equation}
the lowest mass in the above list becomes%
\begin{equation}
M_{2}=8.1TeV.\label{dmp}%
\end{equation}
This becomes a prediction for the mass of a dark matter particle. This is
consistent with the earlier finding of HESS\cite{hess1}. One attractive nature
of the GLMR-RS theory is that the mass ratios listed above are expressed in
terms of fundamental constants in nature.

\section{Evidence for a dark matter particle}

HESS has reported a systematic search for high energy gamma rays from 8
unknown sources\cite{hess2}. Unknown sources could be promising for a dark
matter gamma ray search, since they might be dominated by dark matter, such as
a massive black hole made by the collapse of a predominantly dark matter
object. Such an object might not have the AGN sigature. This is because the
AGN signature requires ordinary atomic matter, while that for unknown sources
may be deficient in ordinary matter. I have reported that the total sum of 8
data samples shows a distinctive peak at 7.6 $\pm$ 0.1 TeV, as shown in Fig.1.
This is consistent with Eq. (\ref{dmp}). This agreement propels further
pursuit of other particles in this model.%
\begin{figure}[ptb]%
\centering
\includegraphics[
height=5.8228in,
width=7.6432in
]%
{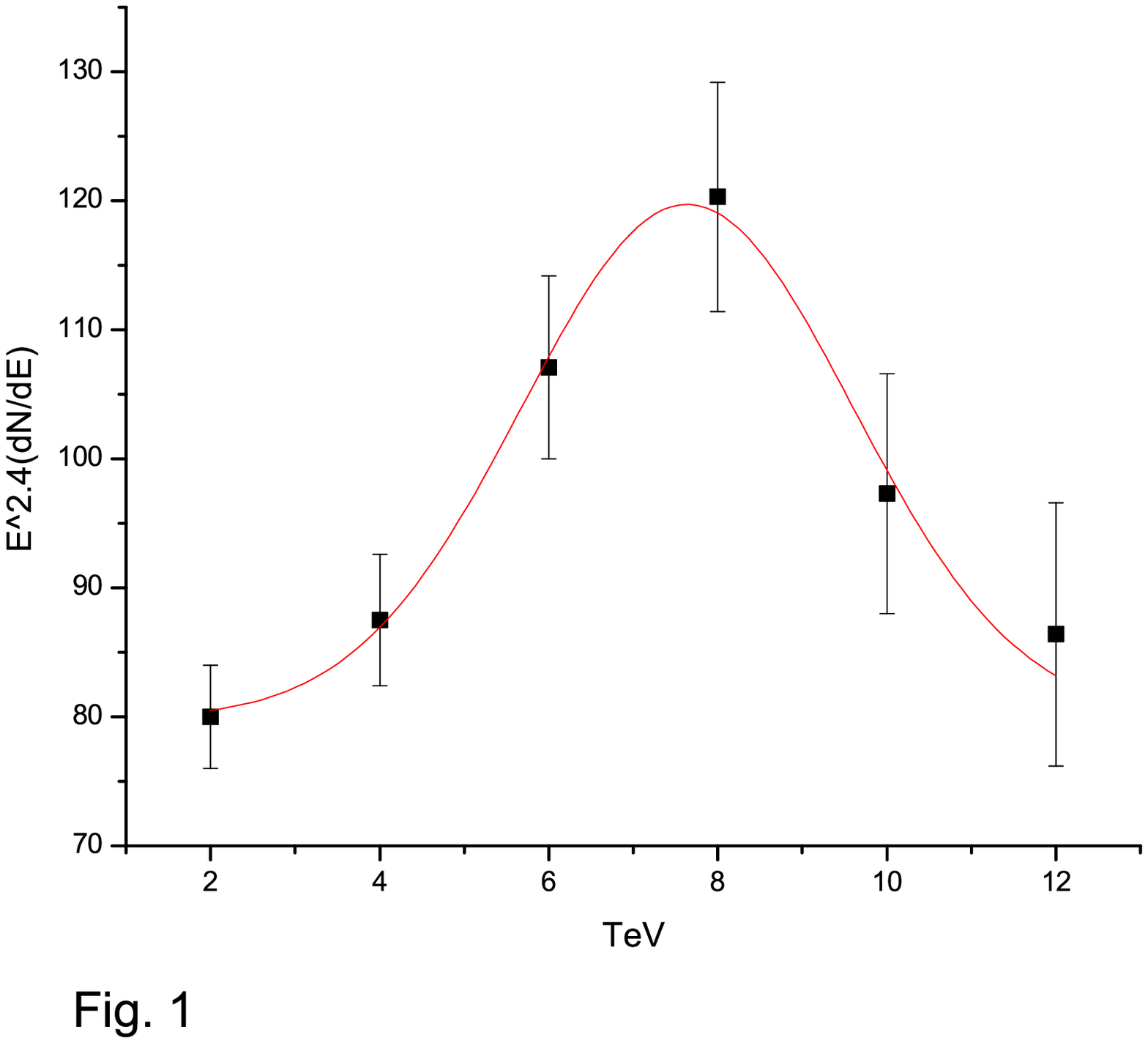}%
\caption{Sum of gamma ray energy spectra of 8 unidentified sources\cite{hess2}%
, in units of 10$^{-12}$(TeV)$^{0.4}$(erg cm$^{-2}$s$^{-1}$). }%
\end{figure}

\section{Signature for unstable members of a dark matter family: A bump in
three body decay}

The family members of the dark matter particle, $M_{2}$, listed above have
masses%
\begin{align}
M_{1}  &  =27TeV,\\
M_{3}  &  =78TeV,
\end{align}

and, of course,%
\begin{equation}
m_{3/2}=3PeV.
\end{equation}
All particles except $M_{2}$ are unstable, as is the case in the standard
model. There are other members in the family at the high end of the mass
range, such as scalars. So the above mass values represent a typical family
member in the lower mass range. Let us use $F$ to denote unstable particles.
They are typically produced as pairs in high energy cosmic ray showers. The
simplest kind of decay is%
\begin{align}
F^{0}-  &  >M_{2}+\mu^{+}+\mu^{-}\\
F^{0}-  &  >M_{2}+\nu+\bar{\nu}\\
F^{0}-  &  >M_{2}+jet+jet
\end{align}
for a neutral component, and%
\begin{equation}
F^{\pm}->M_{2}+\mu^{\pm}+\nu
\end{equation}
for a charged component.

If F pairs are produced at rest in a shower, from the phase space of 3 body
decay one would expect for a decay particle a triangular distribution bounded
by half the parent mass. This can be seen from the gamma ray energy
distribution in 3 gamma decay of orthopositronium, or the electron energy
distribution in muon decay\cite{muon},%
\begin{equation}
\mu->e+\nu+\nu.
\end{equation}
A triangular three body decay distribution is also exhibited in more compact
form in a log-log graph. Since most energy spectra in cosmic ray studies are
exhibited in log-log scale in order to accommodate the large energy range and
power law behavior of back ground processes, this feature is favorable for the
discovery of a three-body-decay bump. Predominantly on the high energy end,
the motion of the parent particle widens the bump, since initial cosmic rays
have a high incident energy. This phenomenon should appear for muons as well
as neutrinos. In other words, the three particles, $M_{1}$, $M_{3}$ and
$m_{3/2}$, would show up as bumps near 13.5 TeV, 39 TeV and 1.5 PeV in the
energy spectra of the muon, neutrino and jet. The only exception would be in
the case where the interaction matrix element suppresses the high energy end
of the spectrum, so that its maxiumum would appear in the middle of the energy
range. This would happen for mu-e decay for V+A theory. If one were to observe
a bump near 1.5 PeV as evidence for $m_{3/2}$, one would know the ratio of the
observed bump with an expected value of half the $m_{3/2}$ mass, 3 PeV.

Fig. 2 of Frejus 94\cite{rhode} shows the energy spectrum for a vertical
stopped muon between 10 TeV and 68 TeV obtained with the Frejus detector. It
clearly suggests that there are enhancements at around 17 TeV and 38 TeV,
consistent with the expectation from bumps\cite{rhode} due to the $M_{1}$ and
$M_{3}$ particles. Definitely one needs more accurate data before drawing a
conclusion. However, it is remarkable that data from two decades ago can
provide such a useful hint. A renewed effort to make an accurate measurement
is encouraged.

Beyond an underground depth of 10,000 kwe, neutrino induced muons start to
dominate, so that energy measurement by stopped muons does not work for the
atmospheric muon spectrum. Therefore, muons beyond 100 TeV cannot be measured
in this manner. In such a case, energy measurement of neutrinos should replace
that of muons for discovering a bump for a dark matter excited state in three
body decay. Alternatively, a different method of measuring muon energy such as
IceCube and Antares should be used.

\section{Muons and neutrinos by neutrino detectors}

Neutrino detectors such as IceCube and Antares can measure muons and neutrinos
at high energy by the Cerenkov method. IceCube has reported a partial spectrum
for high energy muons. A spectrum\cite{icecube1}, \cite{icecube2}, with
average energy 20 TeV has a peak at 10 TeV. This is consistent with a bump at
10-20 TeV, as is suggested by the Frejus data and the $M_{1}$ bump. An IceCube
spectrum with average energy 400 TeV has a peak at 600 TeV. This is consistent
with a bump on a nearby bump at higher energy, such as 1.5 PeV for $m_{3/2}$.
One would like to have IceCube analyze all energy spectra for muons, say at
sea level. As one saw from the previous section, their results might have
important implications for the high energy component of a dark matter family.
A most important task for IceCube and Antares would be to examine whether
bumps similar to those of Frejus 94 can be found in their data. The reason
they didn't examine this process is that they did not realize the significance
of the Frejus bumps.

If the energy spectrum for atmospheric neutrinos were to be measured
accurately in the 10 Tev--10 PeV range, along with that for muons, it might
reveal a dark matter family, and might lead to a new era of particle discovery
through cosmic ray studies.

\section{Production of the knee energy particle, $m_{3/2}$}

In order to produce $m_{3/2\text{ }}$ with a 3 PeV mass, one needs an incident
particle of energy $E$ in the lab frame%
\begin{equation}
E=(3PeV)^{2}/2M
\end{equation}
where $M$ is the target mass. For $M$ = 10 GeV (a Nitrogen nucleus), one needs%
\begin{equation}
E=4.5\ast10^{20}eV, \label{kneeon}%
\end{equation}
i.e., one needs cosmic rays with energy beyond the GZK cuttoff\cite{gzk}. This
may be possible if high energy particles contain dark matter particles. Such a
scenario requires the acceleration of neutral particles. Being a gravitational
acceleration, the model proposed since 1985 by the author does that, and DMPs
are emitted with intensities similar to that of cosmic rays. However, such
intensities are not so high.

If one considers as a target DMPs that reach our neighborhood, one can take
the target mass to be%
\begin{equation}
M=8\text{ }TeV.
\end{equation}
Then the energy required for the production of a knee energy particle of 3 PeV
is%
\begin{equation}
E=5.6\ast10^{17}eV,
\end{equation}
which has a higher intensity than that of the energy in Eq. (\ref{kneeon}). If
one considers dark matter incident particles, then dark matter--dark matter
interactions are supposed to be strong, so that a significant amount of
$m_{3/2}$ particle production can be expected. Also, the distribution of
target dark matter is widely spread beyond the Earth, so that muons and
neutrinos as decay products of unstable $m_{3/2}$ particles may come from a
vast region. It would be a good idea to set up muon or neutrino detectors in a
space station in the future.

If one tries to find a 1.5 PeV peak in the muon spectrum, the lifetime of the
relevant muon becomes 33 seconds from the relativistic effect, so that dark
matter targets within 10$^{12}cm$ can contribute to the process. If one
considers similar events in the neutrino energy spectrum, all dark matter
targets in the whole galaxy can contribute. This implies that the 1.5 PeV peak
in the neutrino energy spectrum should be very large. This phenomenon will
provide decisive evidence for the existence of $m_{3/2}$ at 3 PeV, and at the
same time, it will show the effect of dark matter targets in the whole galaxy.

The same argument can be applied to the production of a $m_{3/2}$ particle by
the collision of dark matter on dark matter, as well as by the collision of
cosmic rays on dark matter. Both processes are strong interactions at this
high energy. The discovery of such a phenomenon is an exciting possibility for
the near future. Definitely, it is worthwhile to see whether a bump at 1.5 PeV
can be observed in present neutrino detectors such as IceCube and Anteras.

\begin{acknowledgments}
\bigskip The author would like to thank David N. Williams for reading the manuscript.
\end{acknowledgments}

\end{document}